\documentclass[doc]{apa7}

\usepackage{enumerate}

\usepackage{tikz, graphicx, booktabs, caption}
\usetikzlibrary{decorations.pathmorphing}
\usepackage{fvextra} 
\VerbatimFootnotes
\usepackage{listings}
\usepackage[american]{babel}
\usepackage{csquotes}

\usepackage[style=apa,backend=biber]{biblatex}
\DeclareLanguageMapping{american}{american-apa}
\hypersetup{colorlinks=true}
\newcommand{\AuxWeb}{\href{https://github.com/sagrawalx/algstat-dif}{GitHub}}

\usepackage{amsmath, amsfonts, calrsfs}
\newcommand{\R}{\mathbb{R}}
\newcommand{\N}{\mathbb{N}}
\newcommand{\cA}{\mathcal{A}}
\newcommand{\cG}{\mathcal{G}}
\newcommand{\cR}{\mathcal{R}}
\newcommand{\cP}{\mathcal{P}}
\newcommand{\h}{\mathrm{no3w}}
\renewcommand{\c}{\mathrm{no23}}
\DeclareMathOperator{\vectorspan}{span}
\DeclareMathOperator{\logit}{logit}

\numberwithin{equation}{section}
\makeatletter
\let\c@figure\c@equation
\let\c@table\c@equation
\makeatother

\title{Using Exact Tests from Algebraic Statistics in Sparse Multi-way Analyses: An Application to Analyzing Differential Item Functioning}
\shorttitle{Using Algebraic Statistics for DIF Analysis}

\authorsnames[1,2,2,2]{Shishir Agrawal, Luis David Garcia Puente, Minho Kim, Flavia Sancier-Barbosa}
\authorsaffiliations{{Department of Mathematics, University of California San Diego}, {Department of Mathematics and Computer Science, Colorado College}}

\authornote{\textbf{Author Order.} We order authors alphabetically.

\textbf{Acknowledgements.} The initial development of this work was funded by the National Science Foundation under Grant LEAPS-MPS-2137661 (PI: Beth Malmskog). We would also like to thank Jane McDougall, David Kahle, Ruriko Yoshida, and Markus Vasquez for helpful conversations.}

\leftheader{S. Agrawal, L. D. Garcia Puente, M. Kim, F. Sancier-Barbosa}

\abstract{Asymptotic goodness-of-fit methods in contingency table analysis can struggle with sparse data, especially in multi-way tables where it can be infeasible to meet sample size requirements for a robust application of distributional assumptions. However, algebraic statistics provides exact alternatives to these classical asymptotic methods that remain viable even with sparse data. We apply these methods to a context in psychometrics and education research that leads naturally to multi-way contingency tables: the analysis of differential item functioning (DIF). We explain concretely how to apply the exact methods of algebraic statistics to DIF analysis using the R package \Verb+algstat+, and we compare their performance to that of classical asymptotic methods.
}


\begin{document}
\maketitle

\section{Introduction}

Contingency table analysis has now been an integral part of statistics and the sciences for well over a century \parencite[cf.][]{Fienberg2007}. Classical methods of analyzing contingency tables, such as Pearson's chi-square test (\citeyear{Pearson1900}), often involve test statistics that asymptotically follow chi-square distributions as sample size increases. Concerns about the applicability of these methods thus arise in small sample size situations, and it was precisely such concerns that led to the development of Fisher's exact test, which assesses independence in $2 \times 2$ tables without relying on large-sample distributional assumptions by conditioning the test on sufficient statistics for the null model \parencites[cf.][]{Fisher1922a}{Fisher1922b}[section 3.5.1]{Agresti2013}. Fisher's test was later generalized to all two-way tables. Although methods of exhaustive enumeration exist \parencites{Mehta1986}{Clarkson1993}, they are frequently untenable and Monte Carlo methods are typically more feasible in practice \parencites{Boyett1979}{Patefield1981}. 

In multi-way tables, many statistical models besides joint independence can be of practical interest (e.g., conditional independence, marginal independence, and homogeneous association). As dimensionality increases, it can become infeasible to meet sample size requirements for applying theoretical asymptotics to test the fit of these models. A temptation may arise to alleviate sample size concerns by decreasing dimensionality via marginalization, but this engenders a risk of erroneous inferences such as the Yule-Simpson paradox \parencites[cf.][]{Yule1903}{Simpson1951}[example 4.1]{Lauritzen1996}. One may resort to collapsing levels, but excessive binning can also be statistically problematic \parencites[cf.][section II.D]{Deckert2019}{Yee2015}.

Exact inference via exhaustive enumeration is almost always untenable in multi-way tables. However, pioneering work of \textcite{Diaconis1998} described a Markov Chain Monte Carlo (MCMC) algorithm for exact inference on multi-way tables by drawing on techniques from algebraic and polyhedral geometry. Their methods apply to \emph{discrete exponential families} \parencites[cf.][]{Sundberg2019}{Efron2023}, which includes a wide range of popular statistical models, including generalized linear models like log-linear models, logistic regressions, and Poisson regressions. Despite early concerns about the computational feasibility of these methods \parencite[e.g.,][88]{Lauritzen1996}, a practical implementation of the methods of Diaconis and Sturmfels now exists in R \parencite{R} in the comprehensive algebraic statistics package \Verb+algstat+ \parencite{algstat}, which builds on a range of software packages including \Verb+4ti2+ \parencite{4ti2} and \Verb+LattE+ \parencites{latte}{latteR}. The field of algebraic statistics has seen significant development since the groundbreaking work of Diaconis and Sturmfels \parencites[e.g.,][]{Chen2006}{Aoki2010}{Dobra2012}{Kahle2018}, and further study continues to be warranted as multivariate datasets with small cell counts proliferate.
 
In order to investigate how the exact methods of algebraic statistics perform against classical asymptotic methods, and to illustrate the use of these methods to practitioners who may have need to analyze sparse multi-way contingency tables, we sought to apply algebraic statistics to a problem in psychometrics and educational research that naturally gives rise to multi-way contingency tables: analyzing \emph{differential item functioning} (DIF, sometimes also called \emph{item bias}). The word \emph{item} in this context refers to a test or survey question, and an item is said to exhibit DIF if there are systematic differences in the distribution of responses from individuals who have the same ability (or some other trait) but who belong to different groups (e.g., different genders, different racial or ethnic groups; cf.\ \cites{Scheuneman1979}{Mellenbergh1982}[section 13.2]{Hambleton1985}). It is typical to treat ability level, group membership, and item response as discrete variables (either categorical or discrete numerical), in which case it becomes necessary to analyze three-way contingency tables. Classical methods of DIF analysis make use of asymptotic goodness-of-fit tests to study these tables. In this study, we propose exact analogs that make use of algebraic statistics. We compare these analogs with their asymptotic counterparts via simulations under various sample sizes and DIF conditions, and demonstrate that these exact methods can allow researchers to better handle smaller sample sizes.

Using DIF analysis as a case study, we aim broadly to bring to light the existence and performance of the exact methods of algebraic statistics for multi-way analyses. By providing practical examples, we hope that researchers who study DIF, as well as other problems that may give rise to sparse multi-way tables, will consider these exact tests as viable alternatives to classical asymptotic tests. 

Our work is organized as follows. In the next section, we give an overview of methods of DIF analysis using two types of statistical models: log-linear models and logistic regressions. We describe classical asymptotic techniques alongside their exact analogs. In the Methods and Results sections, we describe and evaluate simulations that compare the performance of the asymptotic and exact strategies under various sample sizes and DIF conditions. Finally, we discuss our findings and give suggestions for the use of tools from algebraic statistics in DIF analysis and other problems. Code and data related to our work can be found as supplemental material on \AuxWeb. 

\section{Differential Item Functioning}\label{DIF}

Differential item functioning (DIF) occurs when individuals from different groups with the same ability (or another trait) have different probabilities of giving the same response to a test or survey item. Three variables are thus involved in DIF analysis: ability, group, and item response. We assume that response is discrete; items with two possible responses are \emph{dichotomous}, while general items are \emph{polytomous}. As is typical, we will compare across two groups --- a \emph{reference group} and a \emph{focal group} --- though large parts of our discussion apply without difficulty to more than two groups.

An observable variable (e.g., overall test score) is frequently used as a proxy for ability. In order to be able to make use of the exact methods of algebraic statistics for discrete exponential families, we focus on the situation where ability, too, is discretized into finitely many levels. In contexts where ability might naturally be regarded as being continuous, discretization can lead to a loss of information, especially when the number of bins is small \parencites[cf.][]{Yee2015}[section II.D]{Deckert2019}. Nonetheless, binning can also increase parsimony \parencite[section II.C]{Deckert2019} and ``the (cross-sample) stability of the bias analysis'' \parencite[70]{VanDeVijver2021}, and it occurs frequently in tests and surveys that classify individuals based on cutoff scores. Researchers choose bins after observing the overall distribution, with no hard rule for determining cutoffs \parencites[cf.][145]{Scheuneman1979}[138--9]{VanDerFlier1984}[69--70]{VanDeVijver2021}[section 5]{Yee2015}.

A number of parametric formalisms are used for DIF analysis when all three variables of interest are discrete. We focus on two: log-linear models \parencite{Mellenbergh1982} and logistic regressions \parencite{Swaminathan1990}. The two have some differences, and both can be useful in different contexts. Log-linear models may generally be a more flexible option: they require less a priori knowledge about the joint distribution of ability, group, and response, and they can be applied to polytomous items more straightforwardly than logistic regressions \parencite[cf.][]{French1996}. On the other hand, log-linear models treat ability as a nominal variable; logistic regressions treat it as numerical and can thus better retain numerical or ordinal information about ability levels.

Classical methods of DIF analysis rely on the existence of maximum likelihood estimates (MLEs) and on theoretical asymptotics that hold for large samples. We will explain how algebraic statistics provides exact alternatives that do not make use of either MLE existence or asymptotic distributional assumptions. These exact alternatives may thus present some advantages over asymptotic methods in certain situations. In smaller datasets, for example, it is not uncommon for MLEs to fail to exist due to the presence of zero cell counts. If the number of possible responses or ability levels is very large, it may be infeasible to collect samples that are sufficiently large for a sound application of theoretical asymptotics. The possibility of allowing for a larger number of ability levels may also be useful in situations where ability is binned using cutoff values. For example, \textcite[p. 145]{Scheuneman1979} presents several considerations for cutoff selection, one of which is avoiding contingency tables with small counts precisely so that asymptotic methods are viable. Replacing asymptotic methods with exact ones can help remove such considerations from cutoff selection, permitting a more principled approach to binning of ability. 

\subsection{Discrete Models for DIF Analysis} 

Fix a finite set $\cA$ of at least two possible ability levels that a subject can have, and a finite set $\cR$ of at least two possible responses to the test or survey item under study. Let $\cG = \{0, 1\}$ be the set of two groups to which subjects can belong, with $0$ indicating the reference group and $1$ the focal group. Ability $A$, group $G$, and response $R$ are $\cA$-, $\cG$-, and $\cR$-valued random variables, respectively, on the space of respondents. The joint distribution $\pi$ of the triple $(A, G, R)$ is an unknown distribution on $\cA \times \cG \times \cR$.

We start by specifying a \emph{full model}, i.e., a family $\cP$ of distributions on $\cA \times \cG \times \cR$ with no structural zeroes to which $\pi$ is known a priori to belong. In other words, $\cP$ is assumed to lie in the interior of the probability simplex on $\cA \times \cG \times \cR$. Inside $\cP$, there are two nested submodels  
\[ \cP_\c \subseteq \cP_\h \]
that are of interest in DIF analysis. The \emph{conditional independence model} $\cP_\c$ is the submodel in $\cP$ of distributions of $(A, G, R)$ where $G$ and $R$ are conditionally independent given $A$. 
The \emph{homogeneous association model} $\cP_\h$ is the submodel in $\cP$ of distributions of $(A, G, R)$ where $A, G$, and $R$ are homogeneously associated \parencite[section 9.2.2]{Agresti2013}.\footnote{The subscript ``$\c$'' (resp.\ ``$\h$'') refers to the fact that $\cP_\c$ (resp.\ $\cP_\h$) is obtained by intersecting $\cP$ with the hierarchical log-linear model which does not have an interaction between the \emph{second and third} variables (resp.\ which does not have a \emph{three-way} interaction between the variables).}
The item under study \emph{has no DIF} if $\pi \in \cP_\c$. If it does have DIF, it has \emph{uniform DIF} if $\pi \in \cP_\h$ and \emph{nonuniform DIF} otherwise \parencite[cf.][]{Mellenbergh1982}.

Suppose we observe a sample of $n$ subjects in a multinomial sampling scheme, i.e., we make observations of $n$ independent and identically distributed (i.i.d.) random variables with distribution $\pi$. We record these observations in a three-way $\#\cA \times \#\cG \times \#\cR$ contingency table, where the cell corresponding to each $(a, g, r) \in \cA \times \cG \times \cR$ contains the number $n_{a,g,r}$ of subjects in the sample with ability $a$ belonging to group $g$ and responding to the item with response $r$, so that $\sum n_{a,g,r} = n$. Once we settle on strategies for testing the goodness-of-fit of $\cP_\c$ and $\cP_\h$ on the observed data, we can organize these tests into the following two-step \emph{DIF analysis paradigm}.

\begin{enumerate}[(Step 1)]
\item \textbf{DIF Detection.} Test whether $\cP_\c$ fits the observed data. Conclude ``no DIF'' if it fits. Otherwise, proceed to step 2.
\item \textbf{DIF Classification.} Test whether $\cP_\h$ fits the observed data. Conclude ``uniform DIF'' if it fits, and ``nonuniform DIF'' if it doesn't. 
\end{enumerate}

\subsubsection{Strategies for Testing Goodness-of-Fit}

We now describe two possible strategies for testing the goodness-of-fit of the submodels $\cP_*$ appearing above. We call these the ``asymptotic'' and ``exact'' strategies. 

For the asymptotic strategy, we use the observed data to compute a maximum likelihood estimate (MLE) of a distribution in $\cP_*$. This MLE can fail to exist in $\cP_*$, but if it exists, it is usually unique \parencite[cf.][lemma 1.1]{Bogdan2022}. It may sometimes be computed using a closed-form formula, but the computation usually requires numerical approximation. The divergence of the observed data from the MLE can be measured using the likelihood ratio test statistic
\begin{equation} \label{likelihood-ratio} 
G = 2 \sum_{a,g,r} n_{a,g,r} \cdot \ln \left( \frac{n_{a,g,r}}{\hat{\nu}_{a,g,r}} \right), 
\end{equation}
where $\hat{\nu}_{a,g,r}$ are the expected counts under the MLE distribution (which are nonzero because the MLE distribution is in $\cP_*$ and therefore has no structural zeros).\footnote{Pearson's chi-square test statistic $\chi^2 = \sum (n_{a,g,r} - \hat{\nu}_{a,g,r})^2/\hat{\nu}_{a,g,r}$ is a second-order Taylor approximation of $G$ and can be used in its place.}
Wilks's theorem (\citeyear{Wilks1938}; cf.\ \cite[theorem 10.3.3 and section 10.6.2]{Casella2001}) usually guarantees that the distribution of $G$ as sample size increases is asymptotically chi-square with degrees of freedom equal to the codimension of $\cP_*$ inside the probability simplex on $\cA \times \cG \times \cR$. Thus, if the sample size is large enough (a commonly applied heuristic being that at least 80\% of expected counts are at least 5), a goodness-of-fit p-value can be computed as the tail area in the asymptotic chi-square distribution beyond the observed value of $G$.

The asymptotic strategy requires that the MLE for $\cP_*$ exists, but we will see that this can often fail. To account for this in our implementations of asymptotic strategies, we say that the conclusion of the analysis is ``failure'' if the MLE for $\cP_\c$ does not exist. Also, if the MLE for $\cP_\c$ does exist and DIF is detected at Step 1, but the MLE for $\cP_\h$ does not exist, the analysis concludes that the DIF is present but ``unclassifiable.''\footnote{An alternative (and stricter) implementation of the asymptotic strategy might issue ``failure'' and ``unclassifiable'' conclusions when heuristic checks for large enough sample size fail.}

For the exact strategy, we start by fixing sufficient statistics for $n$ i.i.d. observations from $\cP_*$. The \emph{fiber} of the observed table is the (finite) set of all three-way contingency tables with the same value of the sufficient statistics as the observed table. This fiber carries a natural multivariate hypergeometric distribution due to sufficiency of the statistic, and the proportion of tables in the fiber that are at most as likely our observed table is a goodness-of-fit p-value. This is a \emph{conditional p-value} \parencite[section 3.5]{Agresti2013}, because we have conditioned on the sufficient statistics.\footnote{One can also compute a variant conditional p-value by considering the proportion of tables in the fiber whose deviation from the mean table of the fiber is at least as large as that of our observed table.} 

That said, the fiber is frequently too large for exhaustive enumeration, so this p-value can rarely be computed in the way just described. Nonetheless, for a number of common choices of $\cP$, the algebro-geometric techniques\footnotemark{} of \textcite{Diaconis1998} can be used to produce a \emph{Markov basis} for the fiber, i.e., a set of moves between tables which connects any two tables in the fiber.\footnotetext{When $\cP_*$ is a discrete exponential family, its sufficient statistics parametrize a closed toric subvariety of $\operatorname{Spec} k[\cA \times \cG \times \cR]$ over any field $k$ \parencites[section 3.1]{Diaconis1998}[chapter 4]{Sturmfels1996}. The reduced Gr\"obner basis of the ideal of this affine variety (with respect to any monomial ordering) is a finite set of binomials \parencite[corollary 4.4]{Sturmfels1996}, and the exponents that appear in generating binomials correspond to a Markov basis for the fibers of $\cP_*$ \parencite[theorem 3.1]{Diaconis1998}. The problem of computing Markov bases thus reduces to the fundamental problem of computational algebraic geometry: computing Gr\"obner bases. Implicitization \parencites[algorithm 4.5]{Sturmfels1996}[theorem 3.2]{Diaconis1998}[section 3.3]{Cox2015} and integer programming \parencites[algorithm 12.3]{Sturmfels1996}[chapter 1]{Drton2009} can be used to compute Gr\"obner bases of this type.}
This Markov basis can be expensive to compute, but once it has been computed, the Metropolis-Hastings algorithm \parencites{Metropolis1953}{Hastings1970} can generate a Markov chain that randomly samples the fiber, and the proportion of tables in this Markov chain that are at most as likely as our observed table then serves as our goodness-of-fit p-value. Note that the probability of a table in the fiber depends on the cardinality of the fiber (which is frequently large and difficult to compute), but an \emph{unnormalized probability} can be computed directly from the table without knowing the cardinality of the fiber and these unnormalized probabilities can be compared instead.\footnote{The ability to use unnormalized probabilities as a stand-in for true probabilities is also important for implementing the Metropolis-Hastings algorithm.}

\subsubsection{Variants of the DIF Analysis Paradigm}

While we focus on the DIF analysis paradigm as described above, some variants appear in the literature and deserve comment.  All of these variants can be implemented using the exact methods of algebraic statistics. One direction of variation \parencites[cf.][]{Mellenbergh1982}{Swaminathan1990}{Dancer1994} involves switching the order of the goodness-of-fit tests. In other words, we first test the fit of $\cP_\h$, concluding ``nonuniform DIF'' if it doesn't fit. If it does fit, we then test the fit of $\cP_\c$, concluding ``uniform DIF'' if it doesn't fit and ``no DIF'' if it does. Our simulations suggest that this ``swapped'' variant of the paradigm often exhibits worse control over type I error rates for DIF detection, stabilizing noticeably above the nominal rate (cf.\ the full dataset and the secondary comparison plots on \AuxWeb). It can correspondingly have higher power to detect DIF, but we opted against focusing on it because of its inflated type I error rates.

Another direction of variation \parencites[cf.][]{Mellenbergh1982}{Yesiltas2020} involves augmenting the absolute goodness-of-fit test in the DIF classification step with a relative goodness-of-fit test in which we compare the fit of $\cP_\c$ to that of $\cP_\h$, concluding that the item has ``uniform DIF'' only if $\cP_\h$ fits \emph{and} there is a significant difference between the fit of the two models. If $\cP_\h$ fits but the difference in fit is not significant, the DIF is ``unclassifiable.'' This variation makes the classification step more conservative in the sense that it is less likely to conclude ``uniform DIF,'' but the difference was negligible in our simulations: using the asymptotic strategy, this variant led to a different conclusion in only about 0.1\% of simulated tables. We did not include the exact strategy of this variant in our simulations, but a demonstration of conducting model comparisons using algebraic statistics can also be found on \AuxWeb.

A third direction of variation \parencite[cf.][]{Yesiltas2020} becomes particularly relevant in situations where one does not have an a priori guarantee that the true distribution lies in the full model $\cP$. In this situation, the absolute goodness-of-fit test for $\cP_*$ should be replaced with a relative goodness-of-fit test comparing $\cP_*$ to $\cP$. Our simulations generated data according to distributions which do lie in the choices of $\cP$ that we consider, so we did not assess the impact of this direction of variation in this study. 

\subsection{Introducing the HCI Dataset}

To help make our exposition more concrete, we use the \Verb+HCI+ dataset from the R package \Verb+ShinyItemAnalysis+ as an example \parencites{McFarland2017}{Martinkova2018}. The dataset concerns the Homeostasis Concept Inventory (HCI), a 20-item multiple-choice instrument that assesses understanding of the biological concept of homeostasis. Each item has four possible responses, one of which is correct; we treat items as dichotomous, with $R = 0$ indicating any incorrect response and $R = 1$ the correct one. Total scores fall in the 18-point range 3--20, and we discretize these into 6 evenly-spaced ability levels $\cA = \{0, 1, \dotsc, 5\}$. Two possible grouping variables are available in this dataset; here, we consider \Verb+major+, with $G = 1$ indicating subjects intending to major in the life sciences. 

We focus for the time being on item 17, though we return to the remaining items at the end of this section (cf.\ table \ref{hci-table}). Item 17 reads as follows:
\begin{quote}
Baroreceptors sense blood pressure. The baroreceptor nerves are cut so the signal from the baroreceptors is unable to reach the cardiovascular control center. After cutting the nerves, blood pressure will 
\begin{enumerate}[(a)]
\item remain constant.
\item decrease. 
\item increase.
\item become equal to the set-point value.
\end{enumerate}
\end{quote}
The correct answer can be viewed in the supplemental material to \textcite{McFarland2017}. The three-way contingency table for this item is displayed in table \ref{hci-focus-item}. Since all observed cell counts are nonzero, all relevant MLEs will exist for this item. The following code snippet constructs this table \Verb+t+ in R for use further below. 

\begin{table}
\caption{Three-way contingency table for item 17 of the HCI dataset from the R package \texttt{ShinyItemAnalysis} \protect{\parencites{McFarland2017}{Martinkova2018}}. Ability is total score discretized into 6 evenly-spaced bins. $R = 1$ indicates a correct response, and $G = 1$ indicates a student majoring the life sciences.} \label{hci-focus-item}
\centering\footnotesize
\begin{tabular}{ccccc}
\toprule
& \multicolumn{2}{c}{$R = 0$} & \multicolumn{2}{c}{$R = 1$} \\
& $G = 0$ & $G = 1$ & $G = 0$ & $G = 1$ \\
\cmidrule{2-5}
$A = 0$ & 11 & 10 & 4 & 2 \\
$A = 1$ & 32 & 30 & 10 & 7 \\
$A = 2$ & 61 & 66 & 27 & 19 \\
$A = 3$ & 49 & 83 & 14 & 24 \\
$A = 4$ & 35 & 67 & 14 & 47 \\
$A = 5$ & 3 & 11 & 5 & 20 \\ \bottomrule
\end{tabular}
\end{table}

\begin{lstlisting}
library(tidyverse)
library(ShinyItemAnalysis)

# Discretize numerical vector x into discrete levels 0, 1, ..., k-1 
discretize <- \(x, k) as.integer(cut_interval(x, k)) - 1

data(HCI)                                               # Load HCI data
t <- as_tibble(HCI) |> 
     transmute(ability = discretize(total, 6),          # Discretized ability
               group = major,                           # Group by major
               response = `Item 17`) |>                 # Study item 17
     table()                                            # Make three-way table
\end{lstlisting}

\subsection{DIF Analysis with Log-Linear Models}\label{DIFloglinear}

Log-linear models are widely used in categorical data analysis \parencite[cf.][chapter 9]{Agresti2013}. We concentrate on their application to DIF analysis, as introduced by \textcite{Scheuneman1979} and \textcite{Mellenbergh1982}, who focused on the dichotomous case. The polytomous case was treated more fully later \parencites[cf.][]{Dancer1994}{Yesiltas2020}.

We take the full model $\cP$ to be the \emph{saturated model}, i.e., the set of all distributions on $\cA \times \cG \times \cR$ with no structural zeros. The model $\cP$, together with its submodels $\cP_\c \subseteq \cP_\h$, can be described as generalized linear models with a logarithmic link function.\footnote{For more about generalized linear models, see, e.g., \textcite[chapter 4]{Agresti2013}, \textcite[chapter 9]{Sundberg2019}, or \textcite[chapter 3]{Efron2023}.} Namely, for every distribution in $\cP$, we can write
\begin{equation} \label{log-linear-model} 
\ln \Pr[A = a, G = g, R = r] = \lambda + \lambda_{a}^A + \lambda_{g}^G + \lambda_{r}^R + \lambda_{a,g}^{A,G} + \lambda_{a,r}^{A,R} + \lambda_{g,r}^{G,R} + \lambda_{a,g,r}^{A,G,R}
\end{equation}
for all $(a, g, r)$ simultaneously \parencite[equation (9.12)]{Agresti2013}, where the $\lambda$ parameters satisfy appropriate identifiability constraints \parencite[cf.][section 9.1.4]{Agresti2013}. The submodel $\cP_\h$ is the set of the distributions for which 
\[ \ln \Pr[A = a, G = g, R = r] = \lambda + \lambda_{a}^A + \lambda_{g}^G + \lambda_{r}^R + \lambda_{a,g}^{A,G} + \lambda_{a,r}^{A,R} + \lambda_{g,r}^{G,R}, \]
and $\cP_\c$ is the set of the distributions for which 
\[ \ln \Pr[A = a, G = g, R = r] = \lambda + \lambda_{a}^A + \lambda_{g}^G + \lambda_{r}^R + \lambda_{a,g}^{A,G} + \lambda_{a,r}^{A,R}. \]

These models can also be described as discrete exponential families on $\cA \times \cG \times \cR$ of the form studied by \textcite[equation (1.1)]{Diaconis1998}.\footnote{The discrete exponential families of \textcite{Diaconis1998} have trivial \emph{weight function} \parencite[cf.][section 1.1]{Bogdan2022} and the sufficient statistics of a table can be computed as linear functions of cell counts with nonnegative integer coefficients. To help clarify the translation to the notation of \textcite[365]{Diaconis1998}, our lists of quantities that form sufficient statistics are specifying the function $t : \mathcal{X}^N \to \N^d \setminus \{0\}$ that appears there with sample size $N = n$ and $\mathcal{X} = \cA \times \cG \times \cR$. For more about exponential families, see, e.g., \textcites{Sundberg2019}{Bogdan2022}{Efron2023}.} 
Using a subscript $+$ to indicate summing cell counts over all values of that index, the marginals $n_{a,+,r}$ and $n_{a,g,+}$ are sufficient statistics for $\cP_\c$. Adding to these the marginals $n_{+,g,r}$ gives sufficient statistics for $\cP_\h$. All of the individual cell counts $n_{a,g,r}$ form sufficient statistics for $\cP$. 

\subsubsection{Asymptotic Goodness-of-Fit Tests}

There is a closed-form formula for computing the MLE for $\cP_\c$ from the marginals of the observed data; it is the distribution satisfying 
\[ \Pr[A = a, G = g, R = r] = \frac{n_{a,g,+} n_{a,+,r}}{n_{a,+,+}n}, \]
provided that $n_{a,g,+} \neq 0$ and $n_{a,+,r} \neq 0$ for all $(a, g, r)$. The MLE fails to exist if one of these marginals vanishes since we assume no structural zeros \parencite[cf.][corollary 4.9]{Lauritzen1996}. If the MLE exists, the likelihood ratio test statistic \eqref{likelihood-ratio} asymptotically follows a chi-square distribution with $\#\cA (\#\cG - 1) (\#\cR - 1)$ degrees of freedom.  The following R code computes this goodness-of-fit p-value.

\begin{lstlisting}
no23 <- list(c(1, 2), c(1, 3))                          # Facet specs of model
x <- loglin(t, no23, fit = TRUE)                        # Fit the model
pchisq(x$lrt, x$df, lower.tail = FALSE)                 # p-value
\end{lstlisting}

Note that \Verb+loglin+ does not check for MLE existence and may run without complaints even when the MLE does not exist. However, for the table \Verb+t+ from the HCI dataset, this MLE does exist since all observed cell counts are nonzero. Furthermore, 87.5\% of the expected counts under the MLE distribution are at least 5, making it reasonable to use an asymptotic chi-square distribution. The p-value is 0.59, suggesting that the item exhibits no DIF.

Since $\# \cG = 2$, \textcite[proposition 12]{Eriksson2006} give a necessary and sufficient characterization for MLE existence for $\cP_\h$: it is necessary that all of the marginals $n_{a,g,+}, n_{a,+,r}$, and $n_{+,g,r}$ be nonzero, but this is not quite sufficient (in the presence of sampling zeroes).\footnote{The check involves considering all $2 \times 2 \times 2$ collapsings of the table, so it is algorithmically exponential in the dimensions of the table. For other results on MLE existence and uniqueness in discrete exponential families, see \textcite[theorem 9.13]{BarndoffNielsen2014}, \textcite{Jacobson1989}, and \textcite{Bogdan2022}.}
There is no closed-form formula for computing the MLE for $\cP_\h$ from the observed data when it exists, but it can be numerically approximated \parencite[section 9.7]{Agresti2013}. For example, the R function \Verb+loglin+ uses iterative proportional fitting (IPF) to compute this MLE (cf.\ \cite{Haberman1972}). The likelihood ratio test statistic \eqref{likelihood-ratio} asymptotically follows a chi-square distribution with $(\#\cA-1)(\#\cG - 1)(\#\cR - 1)$ degrees of freedom. Though it is not necessary for DIF analysis of the table \Verb+t+ of our running example, an asymptotic goodness-of-fit p-value for $\cP_\h$ can be computed by replacing the facet specification \Verb+no23+ above with \Verb+no3w+, defined as follows.

\begin{lstlisting}
no3w <- list(c(1, 2), c(1, 3), c(2, 3))                 # Facet specs of model
\end{lstlisting}

\subsubsection{Exact Goodness-of-Fit Tests}

Fibers under either model $\cP_*$ are frequently far too large to exhaustively enumerate: the function \Verb+count_tables+ from the R package \Verb+algstat+ \parencite{algstat} can be used to compute fiber cardinalities, and for the table \Verb+t+ of our running example, there are 1,596,426 tables in its fiber under $\cP_\h$, and 103,931,100 under $\cP_\c$. However, the techniques of \textcite{Diaconis1998} can be used to compute Markov chains that randomly sample the fibers. The \Verb+algstat+ function \Verb+loglinear+ implements the relevant functionality. The following uses the same facet specification \Verb+no23+ above and computes the exact goodness-of-fit p-value for $\cP_\c$ and table \Verb+t+ of our running example. 
\begin{lstlisting}
library(algstat)
x <- loglinear(no23, t)                                 # Sample the fiber
x$p.value[["PR"]]                                       # p-value
\end{lstlisting}

We compute a p-value of 0.60,\footnotemark{} suggesting again that item 17 exhibits no DIF.\footnotetext{As the Markov chains vary from run to run, so too do the p-values, but this variation is negligible.}
Replacing \texttt{no23} with \texttt{no3w} above gives a p-value for $\cP_\h$ under the exact strategy, although this is again unnecessary for DIF analysis of this example.

\subsection{DIF Analysis with Logistic Regressions}

Logistic regressions are also widely used in categorical data analysis \parencite[cf.][chapter 5]{Agresti2013}, and \textcite{Swaminathan1990} introduced their application to DIF analysis. We assume throughout our discussion of logistic regression that the item under study is dichotomous, i.e., that $\cR = \{0, 1\}$. Logistic regression treats ability as a discrete numerical variable. In other words, we also assume that $\cA$ is a finite subset of $\R$. 

We start by describing the relevant models as generalized linear models with a logit link function. The full model $\cP$ is the set of distributions on $\cA \times \cG \times \cR$ for which there exist reals $\tau_0, \tau_1, \tau_2, \tau_3$ such that
\begin{equation} \label{logistic-regression-model} 
\logit \Pr[R = 1 \mid A = a, G = g] = \tau_0 + \tau_1a + \tau_2g + \tau_3ag, 
\end{equation}
for all $(a, g)$, where $\logit(x) = \ln(x/(1-x))$. Note that $\cP$ is typically not saturated; it is only saturated when $\#\cA = 2$ (and $\#\cG = 2$). The submodel $\cP_\h$ is the set of distributions satisfying
\[ \logit \Pr[R = 1 \mid A = a, G = g] = \tau_0 + \tau_1a + \tau_2g, \]
and $\cP_\c$ is the set of the distributions satisfying
\[ \logit \Pr[R = 1 \mid A = a, G = g] = \tau_0 + \tau_1a. \]

These models can also be described as discrete exponential families. Sufficient statistics for $\cP_\c$ are the marginals $n_{a,g,+}$ and $n_{+,+,1}$, together with the quantity
\[ \sum_{a \in \cA} a \cdot n_{a,+,1}. \]
Sufficient statistics for $\cP_\h$ are obtained by augmenting the above with the marginal $n_{+,1,1}$. Finally, sufficient statistics for $\cP$ are obtained by augmenting all of the above with
\[ \sum_{a \in \cA} a \cdot n_{a,1,1}. \]

Since the values $a \in \cA$ appear in the formulas for the sufficient statistics, we must assume that $\cA$ is a finite subset of $\N = \{0, 1, \dotsc\}$ in order for this model to be of the form considered by \textcite{Diaconis1998}. However, this assumption is harmless in practice. Indeed, any real number can be approximated to arbitrary precision by a rational, so we can first assume that every $a \in \cA$ is rational, and then we observe that we lose no generality in clearing denominators and translating so that $\cA \subseteq \N$. We will therefore make this harmless assumption on $\cA$ when discussing logistic regressions. 

\subsubsection{Asymptotic Goodness-of-Fit Tests}

The MLE for the logistic regression model $\cP_*$ fails to exist if $n_{a,g,+} = 0$ for some $(a, g) \in \cA \times \cG$, or if there is (either complete or quasicomplete) separation in the data \parencite[theorems 1, 2(i), and 3]{Albert1984}. In general, separation can be detected using linear programming \parencite{Konis2007}, as implemented in the R package \Verb+detectseparation+. However, for the models of interest to us, more concrete characterizations can be used. For $\cP_\c$, separation is equivalent to the existence of $(a^*, r^*) \in \cA \times \cR$ such that  $n_{a,+,r^*} = 0$ for all $a < a^*$ and $n_{a,+,1-r^*} = 0$ for all $a > a^*$. If the MLE exists, the likelihood ratio test statistic \eqref{likelihood-ratio} asymptotically follows a chi-square distribution with $\#\cA \cdot \#\cG - 2$ degrees of freedom \parencite[section 5.2.3]{Agresti2013}. For the table \Verb+t+ of our running example, the MLE exists since all observed cell counts are nonzero. Moreover, 83.3\% of the expected counts under the MLE distribution are at least 5. The R function \Verb+glm+ uses iteratively reweighted least squares (IRLS) to numerically approximate MLEs for logistic regression. The following code computes the p-value for the goodness-of-fit of $\cP_\c$. 

\begin{lstlisting}
u <- as_tibble(t) |>
     mutate(across(everything(), as.integer)) |>
     group_by(ability, group) |> 
     summarize(p = sum(keep(n, response == 1))/sum(n), 
               n = sum(n), 
               .groups = "drop")                        # Transform the data
x <- glm(p ~ ability, binomial, weights = n, data = u)  # Fit no23 model
pchisq(x$deviance, x$df.residual, lower.tail = FALSE)   # p-value
\end{lstlisting}

The p-value is 0.057. With a significance level $\alpha = 0.05$, the resulting DIF analysis suggests that the item has no DIF.

For $\cP_\h$, we can use our assumption that $\cG = \{0,1\}$ to say that separation occurs precisely if there exists  $(a^*_0, a^*_1, r^*) \in \cA \times \cA \times \cR$ such that $n_{a,g,r^*} = 0$ for all $(a, g) \in \cA \times \cG \subseteq \R^2$ strictly to the left of the line $\ell^*$ that passes through $(a_0^*, 0)$ and $(a_1^*, 1)$, and $n_{a, g, 1-r^*} = 0$ for all $(a, g)$ strictly to the right of the same line $\ell^*$. If the MLE exists, the likelihood ratio test statistic asymptotically follows a chi-square distribution with $\#\cA \cdot \#\cG - 3$ degrees of freedom. Replacing the formula \Verb+p ~ ability+ in the first argument of \Verb+glm+ above with \Verb!p ~ ability + group!  gives an asymptotic p-value for the fit of $\cP_\h$ on \Verb+t+, though this is again unnecessary for DIF analysis in this case. 

\subsubsection{Exact Goodness-of-Fit Tests}

The R package \Verb+algstat+ can also be used to compute exact goodness-of-fit p-values for logistic regressions, though, as of this writing, the relevant functionality is not accessible via a single function call (as it is for log-linear models). The following code snippet computes the goodness-of-fit p-value for $\cP_\c$ as follows. We first compute a ``configuration matrix'' \Verb+config_no23+ which specifies the sufficient statistics for $\cP_\c$ in the sense that when the table \Verb+t+ is regarded as the vector \Verb+v+, the product \Verb+config_no23 %*% v+ is a vector of sufficient statistics of \Verb+t+. Then \Verb+markov+ uses this configuration matrix to compute a Markov basis for the fiber, and \Verb+metropolis+ runs the Metropolis-Hastings algorithm to sample the fiber.\footnote{There are 939,003,512,241  tables in the fiber of \Verb+t+ under $\cP_\h$, and  58,866,857,379,038 under $\cP_\c$. To emphasize: the former number is almost 1 trillion, and the latter is larger than 58 trillion! This makes evident that methods of exhaustive enumeration are frequently untenable for multi-way tables.}

\begin{lstlisting}
# Algstat C++ function for computing unnormalized probabilities (UProbs)
computeUProbsCpp <- function(x) {
    .Call("_algstat_computeUProbsCpp", PACKAGE = "algstat", x)
}

d <- map(dimnames(t), as.numeric)                       # Dimension names
A <- t(expand_grid(1, d$ability, d$group))              # Auxiliary matrix
v <- as_tibble(t) |>
    arrange(desc(response), ability, group) |>
    pull(n)                                             # Table as a vector
pr <- computeUProbsCpp(matrix(v))                       # UProb of table

config_no23 <- lawrence(A[1:2, ])                       # Configuration matrix
moves <- markov(config_no23, p = "arb")                 # Markov basis
sample <- metropolis(v, moves, 
                     iter = 10000, 
                     burn = 1000, 
                     thin = 10)$steps                   # Sample fiber
mean(computeUProbsCpp(sample) <= pr)                    # p-value
\end{lstlisting}

For the table \Verb+t+ of our running example, the exact p-value is 0.04. This suggests that $\cP_\c$ does not fit well using a significance level of $\alpha = 0.05$, thus indicating the presence of DIF. To classify the DIF, we compute the p-value for $\cP_\h$ as follows. 
\begin{lstlisting}
config_no3w <- lawrence(A)                              # Configuration matrix
moves <- markov(config_no3w, p = "arb")                 # Markov basis
sample <- metropolis(v, moves,
                     iter = 10000,
                     burn = 1000,
                     thin = 10)$steps                   # Sample fiber
mean(computeUProbsCpp(sample) <= pr)                    # p-value
\end{lstlisting}

The exact p-value is 0.02, suggesting that $\cP_\h$ does not fit well, so we are led to conclude nonuniform DIF. Since we do not have an a priori guarantee that the distribution of the item is in the full logistic regression model $\cP$, the poor fit of $\cP_\c$ and $\cP_\h$ may raise doubt as to whether $\cP$ fits the data. However, the exact p-value for the goodness-of-fit of $\cP$ is 0.12, so the assumption that it fits the data is not unreasonable.

\subsection{Revisiting the HCI Dataset}

We have seen that for item 17, the asymptotic and exact strategies of DIF analysis agree when using log-linear models, but not when using logistic regressions. The results of applying the same four DIF analyses on each item of the HCI dataset are displayed in table \ref{hci-table}. For many of the items, the asymptotic and exact strategies agree. When they differ, the pattern displayed by item 17 is somewhat atypical: differences more frequently result from MLE nonexistence. MLEs for the log-linear conditional independence model fail to exist for 6 of the 20 items when total score is binned into 6 ability levels (and this number increases to 15 out of 20 when 9 bins are used instead). The exact strategy using log-linear models is unaffected by MLE nonexistence and concludes no DIF in every case. 

\begin{table}
\caption{DIF analysis results for each item of the HCI dataset with a significance level of $\alpha = 0.05$. We include results of binning total score into 6 and 9 ability levels. ``Failure'' means that the MLE for $\cP_\c$ does not exist, daggers mark results that change to ``none'' after BH-adjustments for multiple testing for the fit of $\cP_\c$ \protect{\parencite{Benjamini1995}}, and stars mark results where the BH-adjusted p-value for the full model $\cP$ is small (so the assumptions of the DIF analysis may not be satisfied). 
Source code is on \AuxWeb. } \label{hci-table}
\centering\footnotesize
\setlength{\tabcolsep}{6mm}
\begin{tabular}{ccllll}
  \toprule
  & & \multicolumn{2}{l}{Log-Linear Models} & \multicolumn{2}{l}{Logistic Regressions} \\
\cmidrule{3-6}
$\#\cA$ & Item & Asymptotic & Exact & Asymptotic & Exact \\ \midrule
6 &   1 & \emph{failure} & \emph{none} & none & none \\ 
   &   2 & none & none & none & none \\ 
   &   3 & \emph{failure} & \emph{none} & none & none \\ 
   &   4 & none & none & nonuniform\textsuperscript{\textdagger} & nonuniform\textsuperscript{\textdagger} \\ 
   &   5 & none & none & none & none \\ 
   &   6 & none & none & none & none \\ 
   &   7 & none & none & none & none \\ 
   &   8 & \emph{failure} & \emph{none} & none & none \\ 
   &   9 & none & none & none & none \\ 
   &  10 & none & none & nonuniform\textsuperscript{*} & nonuniform\textsuperscript{\textdagger} \\ 
   &  11 & none & none & none & none \\ 
   &  12 & none & none & none & none \\ 
   &  13 & none & none & none & none \\ 
   &  14 & \emph{failure} & \emph{none} & none & none \\ 
   &  15 & none & none & none & none \\ 
   &  16 & none & none & none & none \\ 
   &  17 & none & none & \emph{none} & \emph{nonuniform}\textsuperscript{\textdagger} \\ 
   &  18 & \emph{failure} & \emph{none} & none & none \\ 
   &  19 & \emph{failure} & \emph{none} & none & none \\ 
   &  20 & none & none & none & none \\ 
   \midrule
9 &   1 & \emph{failure} & \emph{none} & none & none \\ 
   &   2 & none & none & none & none \\ 
   &   3 & \emph{failure} & \emph{none} & none & none \\ 
   &   4 & \emph{failure} & \emph{none} & none & none \\ 
   &   5 & \emph{failure} & \emph{none} & none & none \\ 
   &   6 & \emph{failure} & \emph{none} & none & none \\ 
   &   7 & none & none & none & none \\ 
   &   8 & \emph{failure} & \emph{none} & none & none \\ 
   &   9 & \emph{failure} & \emph{none} & none & none \\ 
   &  10 & \emph{failure} & \emph{none} & nonuniform\textsuperscript{\textdagger} & nonuniform\textsuperscript{\textdagger} \\ 
   &  11 & \emph{failure} & \emph{none} & none & none \\ 
   &  12 & \emph{failure} & \emph{none} & none & none \\ 
   &  13 & \emph{failure} & \emph{none} & none & none \\ 
   &  14 & \emph{failure} & \emph{none} & none & none \\ 
   &  15 & none & none & none & none \\ 
   &  16 & none & none & none & none \\ 
   &  17 & none & none & nonuniform\textsuperscript{\textdagger} & nonuniform\textsuperscript{\textdagger} \\ 
   &  18 & \emph{failure} & \emph{none} & none & none \\ 
   &  19 & \emph{failure} & \emph{none} & none & none \\ 
   &  20 & \emph{failure} & \emph{none} & none & none \\ 
   \bottomrule
\end{tabular}

\end{table}

Overall, the results in table \ref{hci-table} are indicative of a fair test, corroborating the claim that ``the HCI assesses understanding of homeostasis for students who are pursuing majors in the life sciences and for students pursuing other majors, although life science majors tend to show the best performance'' \parencite[6]{McFarland2017}.\footnote{The claim that ``life science majors tend to show the best performance'' can be quantified, for example, by tiny p-values for fit of the model of \emph{marginal} independence of ability and group.} 

\section{Methods}\label{Methods}

\subsection{Simulation Models}

To compare the performance of the various DIF analysis methods discussed above, we simulated data using Bock's nominal response model (\citeyear{Bock1972}), which is a nominal polytomous generalization of Rasch's dichotomous model.\footnote{Samejima's graded response model (\citeyear{Samejima1968}) is an \emph{ordinal} generalization of Rasch's dichotomous model, as opposed to Bock's \emph{nominal} one. We use Bock's model because it is simpler to analyze theoretically.} 
Fix a finite set of ability levels $\cA \subseteq \N$, two groups $\cG = \{0, 1\}$, and a finite set of responses $\cR$. Fix also a joint distribution for $(A, G)$. In order to specify the joint distribution of $(A, G, R)$, it is enough to specify the conditional distribution of $R$ given $(A, G)$, which we do by choosing $u, v \in \R^{\cG \times \cR}$ and declaring $\Pr[R = r \mid A = a, G = g]$ to be proportional to $\exp(u(g, r)a + v(g, r))$. Explicitly, 
\begin{equation} \label{bock} 
\Pr[R = r \mid A = a, G = g] = \frac{\exp(u(g, r)a + v(g, r))}{\displaystyle \sum_{r' \in \cR} \exp(u(g, r')a + v(g, r'))}. 
\end{equation}
Since $\cG = \{0, 1\}$, any $x \in \R^{\cG \times \cR}$ corresponds uniquely to $x_0, x_\Delta \in \R^\cR$ such that 
\[ x(g, r) = x_0(r) + x_\Delta(r)g. \]
Decomposing $u$ and $v$ in this way, observe that $(u_0, v_0)$ determines the response distribution in the reference group 0, and $(u_\Delta, v_\Delta)$ describes how the response distribution changes as we go from the reference group 0 to the focal group 1. 

Let $C_\cR$ be the subspace of functions in $\R^{\cG \times \cR}$ which do not depend on the second coordinate, and identify $C_\cR$ with $\R^\cG$. If $u', v' \in C_\cR$, then
\[ \exp((u + u')(g, r)a + (v + v')(g, r)) = \exp(u(g, r)a + v(g, r)) \underbrace{\exp(u'(g)a + v'(g))}_{\text{does not depend on } r}, \]
so the underbraced expression factors out of both the numerator and denominator of \eqref{bock}. In other words, modifying $u$ and $v$ by vectors in $C_\cR$ does not change the resulting distribution, so we lose no generality in choosing $u$ and $v$ to lie in some fixed subspace of $\R^{\cG \times \cR}$ complementary to $C_\cR$.
Since $C_\cR$ can be described as the space of $x \in \R^{\cG \times \cR}$ for which $x_0$ and $x_\Delta$ lie in the subspace $C = \vectorspan\{1\} \subseteq \R^\cR$ of constant functions, we can assume that $u$ and $v$ are chosen so that $u_0, u_\Delta, v_0, v_\Delta \in C^\perp$.

\subsection{Comparison to Log-Linear Models}

The vector $(u_\Delta, v_\Delta)$ determines the qualitative type of DIF in the model. Indeed, observe that
\[ \begin{aligned} \ln &\Pr[A = a, G = g, R = r] \\
&= \ln \Pr[A = a, G = g]\Pr[R = r \mid A = a, G = g] \\
&= \underbrace{\ln \Pr[A = a, G = g] - \ln \left(\sum_{r' \in \cR} \exp(u(g, r')a + v(g, r'))\right)}_{(*)(a, g)} + \ln \exp(u(g, r)a + v(g, r)) \\
&= (*)(a, g) + v_0(r) + u_0(r)a + v_\Delta(r)g + u_\Delta(r)ag.
\end{aligned} \]
The only terms that exhibit dependence on both group and response are the last two, and only the last exhibits dependence on all three variables. Thus $G$ and $R$ are conditionally independent given $A$ if and only if $u_{\Delta} = v_\Delta = 0$, and the three variables are homogeneously associated if and only if $u_{\Delta} = 0$. 

Stated differently, the distribution has no DIF if and only if $u_{\Delta} = v_{\Delta} = 0$, and it has uniform DIF if and only if $u_{\Delta} = 0$ but $v_{\Delta} \neq 0$. In all other cases, the distribution has nonuniform DIF. We call $(u_\Delta, v_\Delta) \in (C^\perp)^2$ the ``DIF vector'' of the model. See figure \ref{fig:dif-vectors}. We use the euclidean norm of $(u_\Delta, v_\Delta)$ as a measure of the size of the DIF.\footnote{It is worth noticing that DIF is a property of the conditional distribution of $R$ given $A$ and $G$. After $\cA$ and $\cR$ are fixed, this conditional distribution is determined by $(u_0, u_\Delta, v_0, v_\Delta) \in (C^\perp)^4$, not by $(u_\Delta, v_\Delta) \in (C^\perp)^2$ alone. There exist measures of DIF size that may be more statistically meaningful than the norm of $(u_\Delta, v_\Delta)$ and which are not functions of that norm alone (cf.\ the Kullback-Leibler heat map on \AuxWeb).}

\begin{figure}
\centering
\begin{tikzpicture}
\tikzstyle{every node}=[font=\footnotesize]
\draw[->, dotted] (-1,0)--(3,0) node[right]{$u_\Delta$};
\draw[->] (0,-1)--(0,3) node[above]{$v_\Delta$};
\fill[black] (0,0) circle (0.75mm);
\node[below=7mm, right=7mm] (T) {No DIF};
\draw[-latex] (T.west) to[left] (2mm,-2mm);
\node[above=17mm, right=7mm] (T) {Uniform DIF};
\draw[-latex] (T.west) to[left] (2mm,17mm);
\end{tikzpicture}
\caption{The space $(C^\perp)^2$ of all DIF vectors $(u_\Delta, v_\Delta)$ for a fixed $(u_0, v_0)$. The picture is heuristic since $\dim(C^\perp) = \#\cR - 1$ in general. The origin corresponds to no DIF, the remainder of the ``vertical axis'' $u_\Delta = 0$ corresponds to uniform DIF, and nonuniform DIF occurs generically everywhere else.} 
\label{fig:dif-vectors}
\end{figure}

\subsection{Comparison to Logistic Regressions}

If $\cR = \{0, 1\}$, then equation \eqref{bock} gives
\[ \begin{aligned} 
\logit &\Pr[R = 1 \mid A = a, G = g] \\
&= u(g, 1)a + v(g, 1) - u(g, 0)a - v(g, 0) \\ 
&= \underbrace{(v_0(1) - v_0(0))}_{\tau_0} + \underbrace{(u_0(1) - u_0(0))}_{\tau_1}a + \underbrace{(v_\Delta(1) - v_\Delta(0))}_{\tau_2}g + \underbrace{(u_\Delta(1) - u_\Delta(0))}_{\tau_3}ag. 
\end{aligned} \]
In other words, the dichotomous items we simulate using model \eqref{bock} induce distributions on $\cA \times \cG \times \cR$ that lie in the full logistic regression model \eqref{logistic-regression-model}.

\subsection{Simulations and Assessment}

We chose 16 integer seeds to create a variety of simulation models. Using each of these seeds, we generate a random choice of $\cA$ and $\cR$, a joint distribution of $(A, G)$, a vector $(u_0, v_0) \in (C^\perp)^2$, and five DIF vectors $(u_\Delta, v_\Delta) \in (C^\perp)^2$. The five DIF vectors are:
\begin{enumerate}[(1)] 
\item the zero vector,
\item a uniform unit vector (i.e., with $u_\Delta = 0$), 
\item twice the same uniform vector, 
\item a nonuniform (i.e., generic) unit vector, and 
\item twice the same nonuniform vector. 
\end{enumerate}
In other words, each integer seed gives rise to five joint distributions of $(A, G, R)$, all of which have the same distribution of $(A, G)$, and the same conditional distribution of $R$ given $A$ and $G = 0$, but which exhibit five different types and magnitudes of DIF. Model parameters generated by 3 of the 16 integer seeds that we used are displayed in table \ref{dif-configurations}. For model parameters for other seeds, see \AuxWeb.

For each of the five joint distributions of $(A, G, R)$ corresponding to each seed, we simulate collecting samples of various sizes. Each sample is handed off to a number of DIF analysis methods: every table is handed off to the asymptotic and exact versions of the method using log-linear models, and every dichotomous table is also handed off to the asymptotic and exact versions of the method using logistic regression. The result of each method can then be compared against the known ``true'' value of the type of DIF in the simulation model. This allows us to compare the performance of the DIF analysis methods as sample sizes and DIF conditions vary.

\begin{table}
\caption{Simulation model parameters generated by 3 of the 16 integer seeds we chose. In all instances, $\cG = \{0,1\}$. These are nominal response models, so elements of $\cR$ are just labels. The displayed DIF vectors $(u_\Delta, v_\Delta)$ have norm 1 (multiply by 2 for norm 2).} \label{dif-configurations}
\centering\footnotesize
\begin{tabular}{lll}
  \toprule
Seed & Parameter & Value \\ 
  \midrule
1930 & $\Pr[G = 0]$ & 0.366 \\ 
   & $\cA$ & \{0, 1, 2, 3, 4, 5, 6\} \\ 
   & $\Pr[A \mid G = 0]$ & (0.002, 0.020, 0.095, 0.235, 0.328, 0.244, 0.076) \\ 
   & $\Pr[A \mid G = 1]$ & (0.005, 0.043, 0.152, 0.286, 0.302, 0.170, 0.040) \\ 
   & $\cR$ & \{0, 1\} \\ 
   & $(u_0, v_0)$ & (0.407, -0.407, 0.579, -0.579) \\ 
   & Uniform $(u_\Delta, v_\Delta)$ & (0.000, 0.000, 0.707, -0.707) \\ 
   & Nonuniform $(u_\Delta, v_\Delta)$ & (0.004, -0.004, 0.707, -0.707) \\ 
   \midrule
1947 & $\Pr[G = 0]$ & 0.429 \\ 
   & $\cA$ & \{0, 1\} \\ 
   & $\Pr[A \mid G = 0]$ & (0.368, 0.632) \\ 
   & $\Pr[A \mid G = 1]$ & (0.203, 0.797) \\ 
   & $\cR$ & \{0, 1\} \\ 
   & $(u_0, v_0)$ & (-0.331, 0.331, -0.625, 0.625) \\ 
   & Uniform $(u_\Delta, v_\Delta)$ & (0.000, 0.000, 0.707, -0.707) \\ 
   & Nonuniform $(u_\Delta, v_\Delta)$ & (-0.539, 0.539, 0.458, -0.458) \\ 
   \midrule
1948 & $\Pr[G = 0]$ & 0.514 \\ 
   & $\cA$ & \{0, 1, 2, 3\} \\ 
   & $\Pr[A \mid G = 0]$ & (0.091, 0.335, 0.408, 0.166) \\ 
   & $\Pr[A \mid G = 1]$ & (0.037, 0.222, 0.444, 0.296) \\ 
   & $\cR$ & \{0, 1, 2\} \\ 
   & $(u_0, v_0)$ & (0.187, -0.690, 0.503, -0.269, -0.118, 0.387) \\ 
   & Uniform $(u_\Delta, v_\Delta)$ & (0.000, 0.000, 0.000, -0.455, -0.360, 0.815) \\ 
   & Nonuniform $(u_\Delta, v_\Delta)$ & (0.207, -0.009, -0.198, 0.752, -0.188, -0.563) \\ 
   \bottomrule
\end{tabular}

\end{table}
	
\section{Results}\label{Results}

\begin{figure}
\centering
\includegraphics[width=\textwidth]{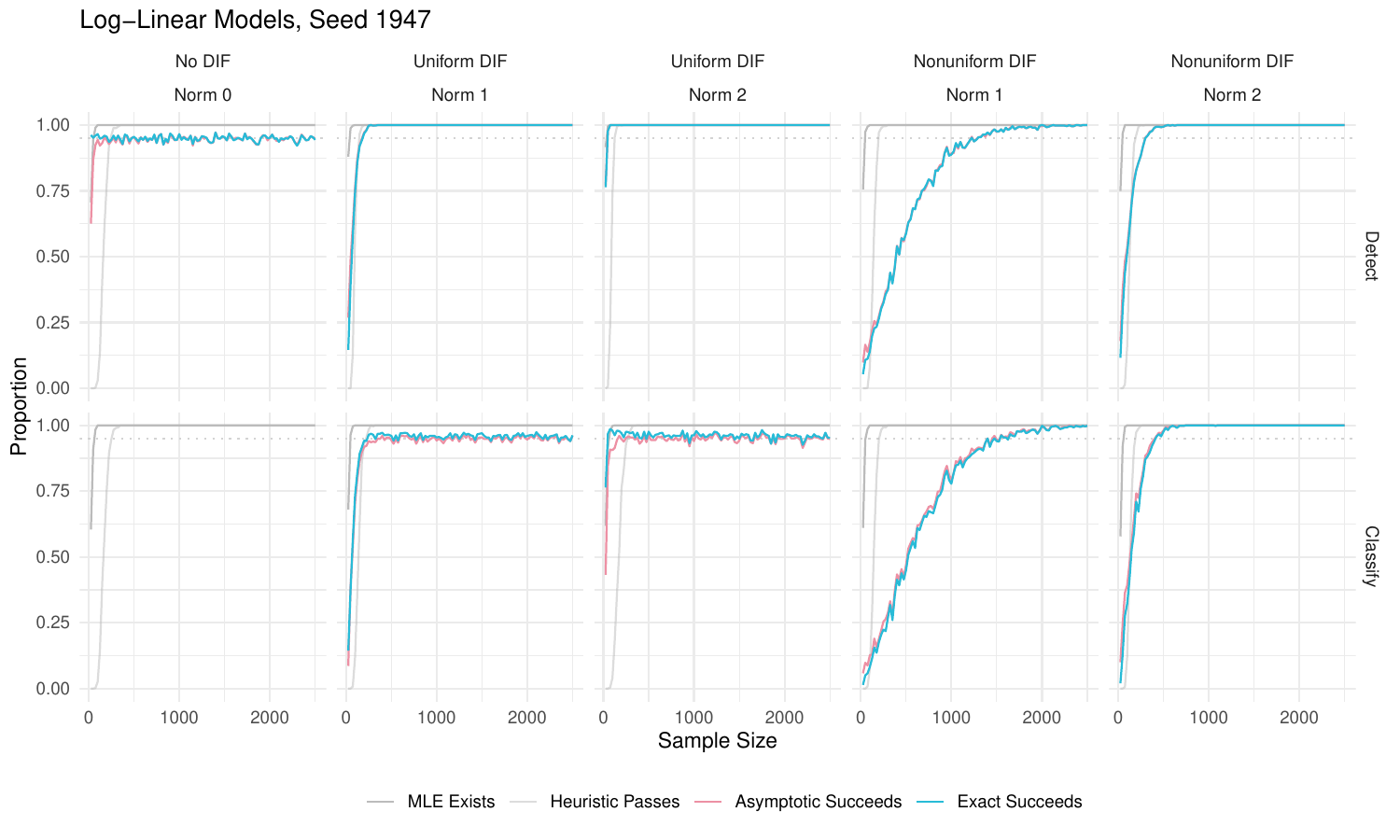}
\caption{Results of DIF analysis using log-linear models on $2 \times 2 \times 2$ tables. The asymptotic and exact strategies converge rapidly in performance.}
\label{fig:llm-fast}
\end{figure}

\begin{figure}
\centering
\includegraphics[width=\textwidth]{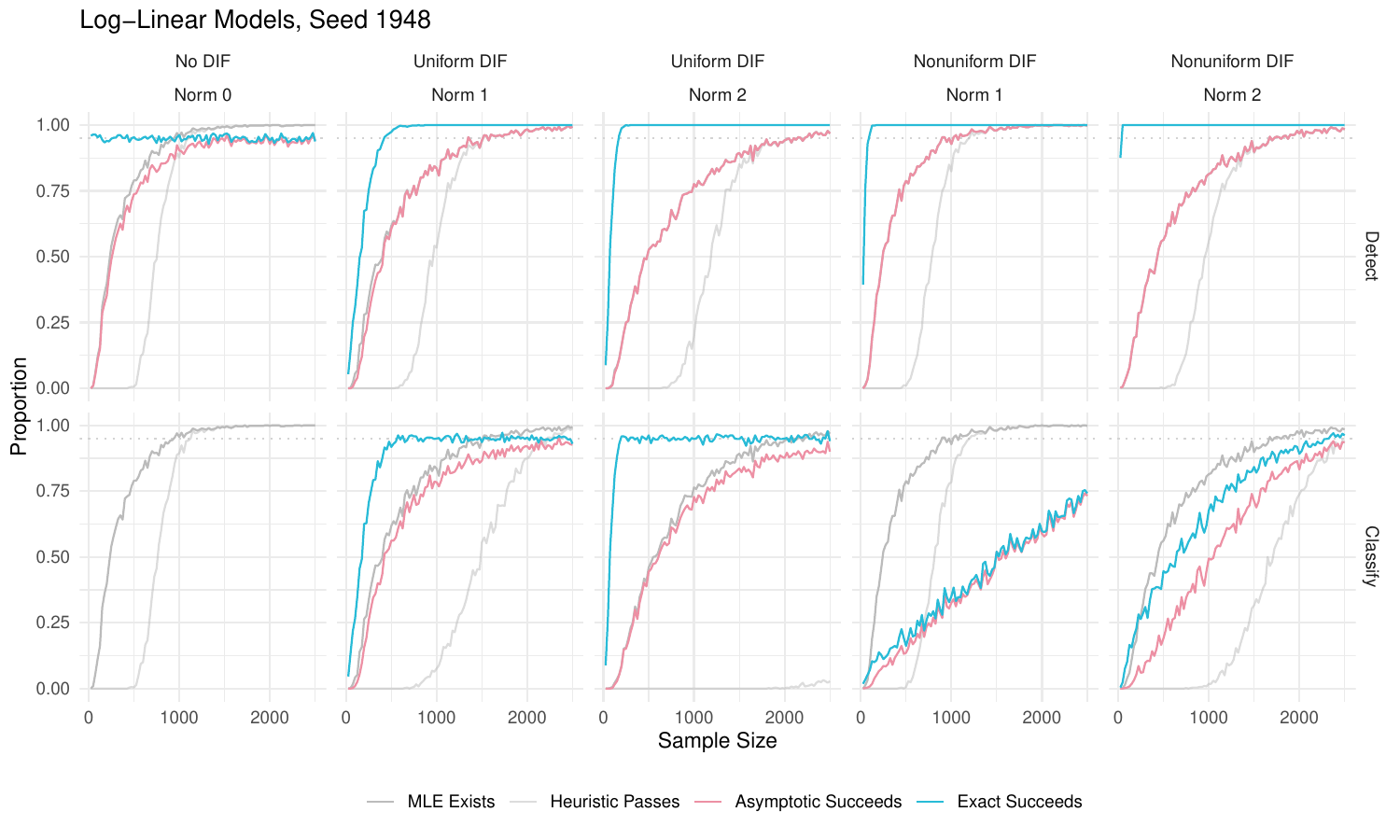}
\caption{Results of DIF analysis using log-linear models on $4 \times 2 \times 3$ tables. There are sizable gaps in performance between the asymptotic and exact strategies.}
\label{fig:llm-slow}
\end{figure}

\begin{figure}
\centering
\includegraphics[width=\textwidth]{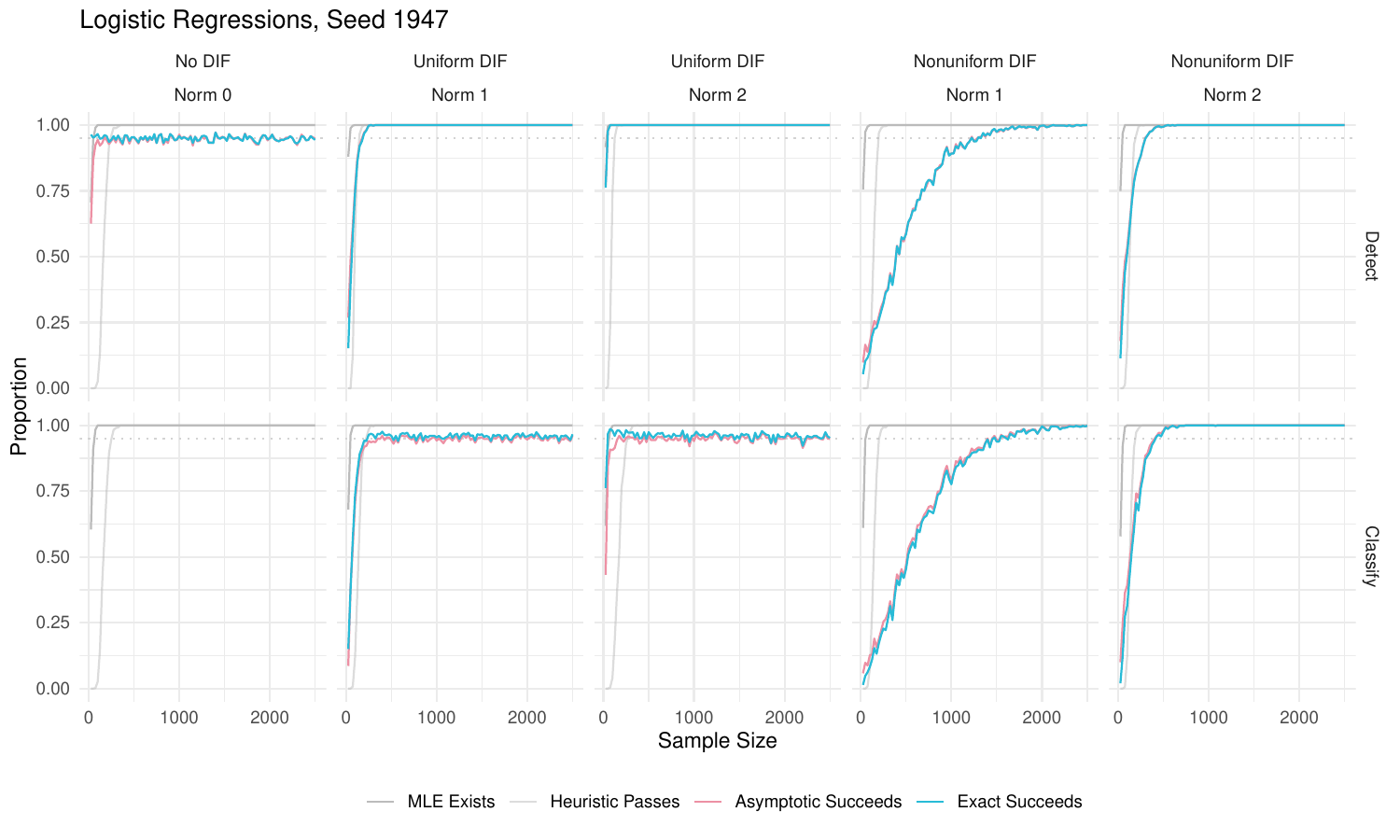}
\caption{Results of DIF analysis using logistic regressions on $2 \times 2 \times 2$ tables. As in figure \ref{fig:llm-fast}, the asymptotic and exact strategies converge rapidly in performance.}
\label{fig:lrm-fast}
\end{figure}

\begin{figure}
\centering
\includegraphics[width=\textwidth]{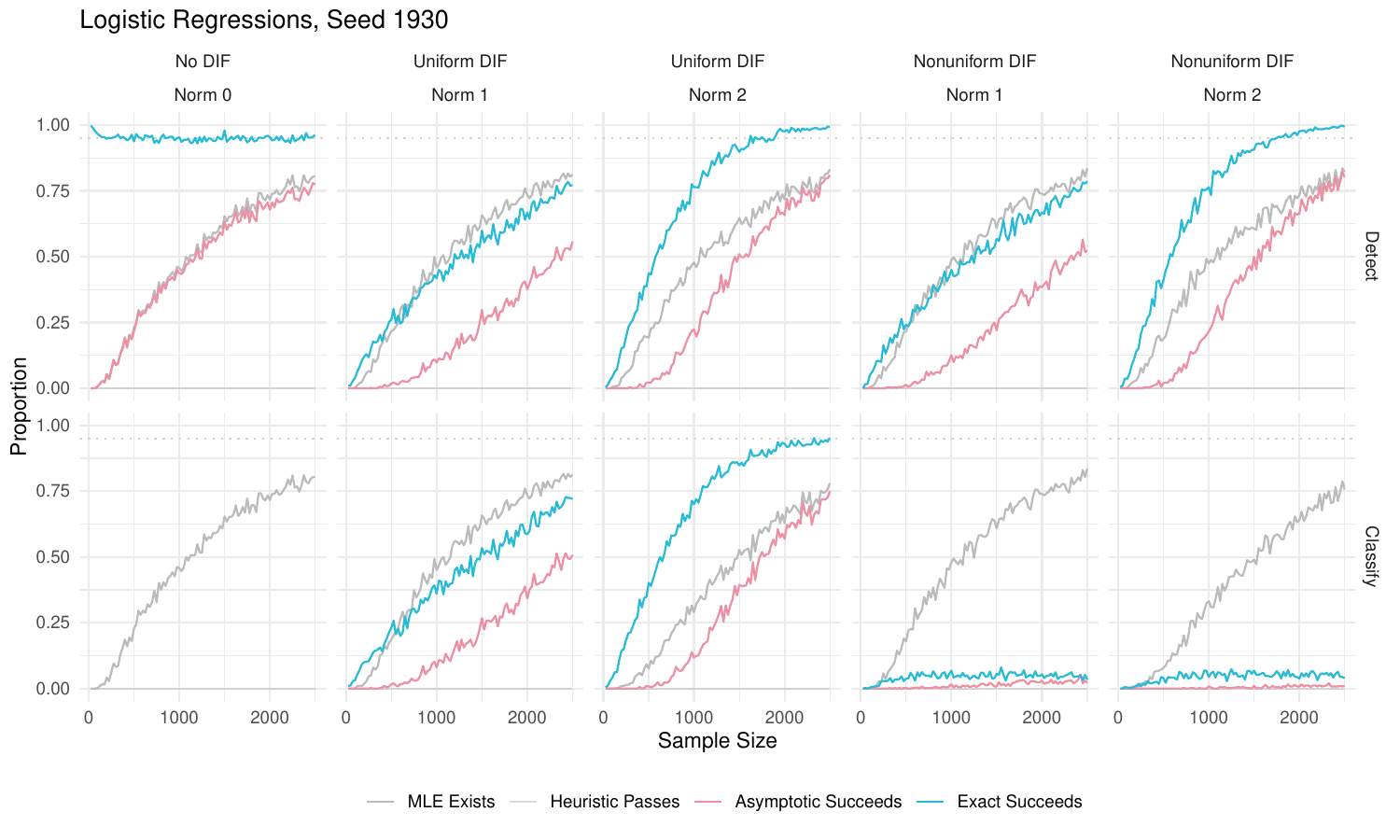}
\caption{Results of DIF analysis using logistic regressions on $7 \times 2 \times 2$ tables. There are sizable gaps in performance between the asymptotic and exact strategies, as in figure \ref{fig:llm-slow}. Note that classification of nonuniform DIF is extremely difficult for both the algebraic and asymptotic methods because the nonuniform DIF vector corresponding to this seed has $u_\Delta$ very close to 0 (cf.\ table \ref{dif-configurations} and the Kullback-Leibler heat map on \AuxWeb).}
\label{fig:lrm-slow}
\end{figure}

Representative simulation results for the 3 integer seeds appearing in table \ref{dif-configurations} are displayed in figures \ref{fig:llm-fast} through \ref{fig:lrm-slow}. The figures are structured as follows. The first row indicates results related to DIF detection, and the second to DIF classification. Gray lines in each plot indicate rates of MLE existence; these upper bound the performance of the asymptotic method, since the asymptotic methods issue ``failure'' or ``unclassifiable'' conclusions when MLEs fail to exist. The first row indicates MLE existence for $\cP_\c$, and the second for $\cP_\h$. (Also indicated, in a lighter gray, are rates of meeting the heuristic for applying asymptotic tests which requires that at least 80\% of expected counts under the MLE distribution in $\cP_*$ be at least 5.) The colored lines in the first plot shows the proportion of tables generated using a no DIF model that were correctly classified as having no DIF (i.e., roughly one minus type I error rate for DIF detection). The colored lines in the remaining four plots of the first row indicate power for DIF detection. In the second row, colored lines in the second and third plots show the proportion of tables generated using a uniform DIF model that were correctly classified as such (i.e., roughly one minus type I error for DIF classification), and in the fourth and fifth plots, they indicate power for DIF classification. 

Figures \ref{fig:llm-fast} and \ref{fig:llm-slow} depict results of DIF analyses using log-linear models. Figure \ref{fig:llm-fast} shows results from a model of low dimensionality, where tables are $2 \times 2 \times 2$. MLEs frequently exist in these tables (except for the smallest of sample sizes), and the performance of asymptotic and exact methods converge quickly to each other. Type I error rates for both detection and classification converge quickly to the nominal rate of $\alpha = 0.05$. Power for DIF detection converges quickly to 100\%, though power for classification takes longer. 
Figure \ref{fig:llm-slow} shows analogous results from a model with higher dimensionality, where tables are $4 \times 2 \times 3$. The difference between the two strategies is stark. The exact strategy holds type I error rates at nominal values and obtains substantial power even in small sample situations where MLEs are extremely unlikely to exist. In fact, in this situation MLEs frequently fail to exist even for the largest samples we generated, and the two methods do not converge in performance until the very largest sample sizes in our simulation. 

Figures \ref{fig:lrm-fast} and \ref{fig:lrm-slow} depict results of DIF analyses using logistic regression models. They exhibit roughly the same patterns as the ones for log-linear models.

Overall, our simulations demonstrate that the exact strategies of algebraic statistics are very consistent at holding type I error rates at nominal values. Conditional p-values are known to exhibit type I error rates slightly below nominal due to discreteness of the distribution \parencite[section 3.5.5]{Agresti2013}. This conservatism is visible in our simulations, but it is minimal: type I error rates for the exact methods converge rapidly to nominal. In the range of sample sizes we considered, most of the discrepancy between the exact and asymptotic strategies arises due to MLE nonexistence, which is especially prevalent for smaller sample sizes and many-celled tables, although not restricted to such cases. It is not infrequent to observe enough sampling zeros that MLEs fail to exist, even with $4 \times 2 \times 3$ tables as in figure \ref{fig:llm-slow} or $7 \times 2 \times 2$ tables as in figure \ref{fig:lrm-slow}, and the problem would only grow as table dimensionality increases. This bottlenecks the performance of asymptotic methods, but the exact methods of algebraic statistics are unaffected: they remain viable with competitive power and type I error control. When MLE existence is not an issue, there are typically only minor discrepancies between the two strategies. 

\section{Discussion}\label{Discussion}

Our study has sought to demonstrate that the exact methods of algebraic statistics, and the practical implementation of these methods that is available through the R package \verb+algstat+, may prove to be a useful tool to researchers working in psychometrics,  education research, and other fields who find themselves faced with analyzing sparse multi-way contingency tables. These techniques can be used even when MLEs fail to exist, and can help circumvent the need to collapse levels in order to meet sample size requirements for applying asymptotics. 

The techniques apply broadly, not only to the log-linear models and logistic regressions we have described above. For example, there is another method of DIF analysis involving the Cochran-Mantel-Haenszel (CMH) procedure which fits squarely into the framework we have described. For this method, one takes $\cR = \{0,1\}$ and $\cP$ to be the set of distributions satisfying $\logit \Pr[R = 1 \mid A = a, G = g] = \tau_1(a) + \tau_2 g$ for $\tau_1 \in \R^\cA$ and $\tau_2 \in \R$ \parencite[cf.][363--4]{Swaminathan1990}. Then $\cP_\h = \cP$, so one cannot use this model for DIF classification, but $\cP_\c$ is the proper submodel of distributions where $\tau_2 = 0$, so it is still possible to use this family for DIF detection. Goodness-of-fit for the CMH procedure is typically assessed through asymptotics, but exact tests can also be used since these families are discrete exponential families of the form studied by \textcite[equation (1.1)]{Diaconis1998}: the marginals $n_{a,g,+}$ and $n_{a,+,1}$ are sufficient statistics for $\cP_\c$, and this list is augmented with $n_{+,1,1}$ for $\cP_\h = \cP$.

Similarly, Bock's model \eqref{bock} is also a discrete exponential family of the form studied by \textcite[equation (1.1)]{Diaconis1998}.\footnote{In their notation, the map $T : \cA \times \cG \times \cR \to (\R^{\cG \times \cR})^2 = (\R^2)^{\cG \times \cR}$ is given by $T(a, g, r) = [(g, r) \mapsto (a, 1)]$.} Thus assessments of the fit of this model $\cP$, and of its submodels $\cP_\c \subseteq \cP_\h$, can also be used to detect and classify DIF. Asymptotic strategies for assessing this fit would rely on first computing an MLE, which can be done using methods described by \textcite{Bock1972}, \textcite[chapter 3]{Sundberg2019}, or \textcite[section 2.6]{Efron2023}. Alternatively, one can employ the exact strategy of \textcite{Diaconis1998} to assess the fit, following the pattern we have described in this study. 

In fact, goodness-of-fit tests for discrete exponential families are applied in item response theory beyond just DIF analysis \parencites[cf.][chapters 8--9]{Hambleton1985}{Kelderman1996}[chapter 5]{Ostini2006}. When data is sparse, these tests for model fit may also benefit from the techniques of algebraic statistics. 

Our investigations suggest some natural avenues for future research. For example, our simulations capped the number of ability and response levels at 7, and in this range, the relevant Markov basis computations were fairly quick. Markov basis computations for the $9 \times 2 \times 2$ tables that occurred in table \ref{hci-table} were still quite feasible for a personal computer. That said, computation of Markov bases is known generally to be a bottleneck for applying the exact techniques of algebraic statistics. However, a number of techniques have been proposed in the literature to attempt to bypass this bottleneck. For example, \textcite[chapter 16]{Aoki2010} describes a technique in which Markov bases are replaced by lattice bases, which are easier to compute but require more care when generating the Markov chain. Alternatively, \textcite{Dobra2012} describes a technique where Markov bases are computed dynamically at each step of the Markov chain. Yet other techniques are described by \textcite{Chen2006}, \textcite{Chen2010} and \textcite{Rapallo2010}. It would be interesting to investigate applications of these techniques to DIF analysis and item response theory, to other areas in psychometrics and education research, and to the sciences more broadly.

\printbibliography

\end{document}